\documentclass[11pt,a4paper]{JHEP3}
\usepackage{epsfig,amssymb,euscript,bbold}
\usepackage{amsmath}

\allowdisplaybreaks
\preprint{DAMTP-2007-36}

\def\be{\begin{equation}}
\def\ee{\end{equation}}
\def\bea{\begin{eqnarray}}
\def\eea{\end{eqnarray}}

\def\BHG{{\bf {\rm BHG}}}
\def\tBHG{\widehat {\BHG}}

\numberwithin{equation}{section}

\title{Black Hole Giants}
\author{Aninda Sinha and Julian Sonner  \\
Department of Applied Mathematics and Theoretical Physics, \\
University of Cambridge, UK\\
{\tt a.sinha, j.sonner@damtp.cam.ac.uk}}
\abstract{We investigate giant and dual giant
    type BPS configurations in the near-horizon geometry of a certain
    $\frac{1}{16}$-BPS AdS$_5$ black hole. By quantising the space of
    solutions we count the dual giant configurations and compare with
    the black hole entropy. This suggests a missing degeneracy factor
    which we argue comes from an angular momentum quantum number. From
    the D-brane world volume this arises from BPS electromagnetic
    waves. We study these waves in the context of giants and dual
    giants in the black hole near-horizon geometry. We further
    demonstrate that turning on waves on the world-volume of
    $\tfrac{1}{8}$-BPS dual giants in AdS$_5\times$ S$^5$ leads to $\frac{1}{16}$-BPS
    states with an additional angular momentum quantum number.}
\begin{document}
\section{Introduction}

String theory has been successful in accounting for the statistical
entropy of many supersymmetric asymptotically flat black holes
\cite{sv,rev}. Three years ago Gutowski and Reall discovered
supersymmetric asymptotically AdS$_5$ black holes with regular
horizons \cite{gr1,gr2,cclp1,cclp2,klr, hed}. A microscopic
understanding of these black holes is an important open problem in AdS/CFT.

The simplest such asymptotically AdS$_5$ black hole rotates with equal
angular momenta in two orthogonal planes in AdS$_5$ directions and carries a single $U(1)$
electric charge. The entropy of this black hole is known to be
\be
\label{entropy}
S_{BH}={\pi^2\over 2 G_5 }\omega^3 \sqrt{1+{3\omega^2\over
    4 l^2}}\,,
\ee
where $\omega$ is a parameter related to the black hole angular
momentum and electric charge and $l$ is the AdS$_5$ radius.  As was
shown in \cite{ggs}, when lifted to a 10-dimensional type IIB
solution, the geometry asymptotes to AdS$_5 \times $S$^5$ and preserves
just two supersymmetries. Since only the five-form flux is turned on, the
microstates of this black hole may be thought to be some configuration
of multiple giant gravitons \cite{Grisaru:2000zn,McGreevy:2000cw},  which preserve $\frac{1}{16}$ of the
supersymmetries of AdS$_5 \times$ S$^5$. The construction and counting
of such states is proving to be a difficult, and as yet unsolved,
problem (for some related progress see
\cite{kl,Feng:2007ur,kmmr,bglm,Silva:2006xv,Dias:2007dj}). So it is natural to look for other avenues to address the
problem of microstate counting for these black holes. For instance, a Fermi surface model was proposed for 
a microscopic description for these black holes in \cite{brs} where a qualitative agreement was found. 

Around two years ago, Strominger and collaborators provided a specific
example of a four-charge black hole carrying D0 and D4 charges with near-horizon geometry AdS$_2\times$ S$^2\times$ CY$_3$ where near-horizon
microstates could account for the entropy \cite{gssy,gsy}. The
microstates involved in this derivation did not preserve any of the
asymptotic supersymmetries. One reason for this somewhat surprising
feature is that supersymmetric quantum mechanics tells us that the
microstates preserving the asymptotic supersymmetries are
non-normalisable \cite{fr} and hence should not be included in the counting. The
way out of this conundrum was to transform to global time \cite{Claus:1998ts} and use
eigenstates of the global Hamiltonian to do the counting. In Poincar\'e
time, these states corresponded to D0 brane states popping in and out
of the horizon.

Motivated by this picture, the near-horizon geometry of the
simplest Gutowski-Reall AdS$_5$ 
black hole was studied in some detail in \cite{sss}. There it was shown that there
is a doubling of supersymmetries near the horizon. The superisometry
group of the horizon was found to be $SU(1,1|1)$.  When lifted to ten
dimensions, the near-horizon geometry has a deformed three-sphere $\tilde{\text{S}}^3$ and a deformed five-sphere $\tilde{\text{S}}^5$ with a fibration of the time
coordinate of AdS$_2$ over them. The AdS$_2$ part of the geometry can
be written in both global and Poincare coordinates. We will call
D3-branes wrapping three of the $\tilde{\text{S}}^5$ directions, black hole giant
gravitons ($\BHG$) while D3-branes wrapping the $\tilde{\text{S}}^3$ will be called
black hole dual giants ($\tBHG$). It was shown in \cite{sss} that giant and dual giant type
probes which preserve half the near-horizon supersymmetries exist in
the lifted geometry. In AdS$_2$ Poincar\'e coordinates, the probes have zero energy and
preserve exactly the asymptotic supersymmetries. In AdS$_2$ global coordinates, the probes have 
non-trivial Hamiltonians and preserve none of the asymptotic supersymmetries. In this case
both $\BHG$ and $\tBHG$ preserve the same fraction of the near-horizon supersymmetries.
One naturally wonders if these near-horizon
microstates could be used to account for the microscopic entropy of the
black hole. Another reason to expect this to be the case is that the conserved charges
of the black hole can be extracted completely from the near-horizon geometry
as was shown in \cite{Suryanarayana:2007rk}.

In this paper we quantise the phase space of solutions of the $\tBHG$s
in AdS$_2$ global coordinates
and count them. We find that there is an exponential degeneracy and
hence a large contribution to the microstates from these
solutions. The leading order result is off by a degeneracy factor
which we argue is the result of a missing quantum number. 


Motivated by the missing quantum number we study world-volume fluxes
which preserve the same supersymmetry as the original solutions. We
find that a whole class of solutions exist where electromagnetic waves
can be turned on in the fibre direction after writing the deformed
3-spheres as Hopf fibrations over $\mathbb{C}P^1$. These waves contribute to
the missing angular momentum quantum number. The resulting equations
of motion are very similar to the $\frac{1}{8}$-BPS AdS$_5\times$ S$^5$ giants with fluxes
which were studied in \cite{Kim:2005mw}. We will demonstrate that
turning on world-volume fluxes on $\frac{1}{8}$-BPS AdS$_5\times $S$^5$ dual
giants will generically break supersymmetry by a further half. This
can be anticipated by noting that the most general $\frac{1}{8}$-BPS dual giant
configuration \cite{ms} is known to be spherically symmetric and
turning on waves will generically break this spherical symmetry. We
provide a simple maximisation argument motivated by \cite{kmmr} to
show how the near-horizon and asymptotic states could be used to
account for the macroscopic entropy. The direct way of doing this is
by quantising the new phase space which we have not attempted in this paper.

In order to account for the full black hole entropy, one possibly
needs to turn on mechanical waves on the world-volume as well. We will
not have anything to say about these but will leave this as an open
problem. In the final solution to this problem from near-horizon
microstates, we feel our BPS analysis of world-volume electromagnetic
waves will be important. Our analysis may also be helpful in
developing an understanding of how the black hole superconformal
quantum mechanics is embedded in ${\cal N}=4$ super Yang Mills.

The paper is organised as follows. In section 2, we review the near
horizon geometry and probes of the black hole under investigation. In
section 3, we count dual giant type configurations and motivate the
addition of fluxes on the world-volume. In section 4, we study near
horizon giant and dual giant type configurations with world-volume
fluxes which preserve the same supersymmetry as those without
fluxes. We conclude with a discussion and some speculative comments in
section 5. Calculational details of the supersymmetry analysis are given in appendices A and B.

\section{Review of the Near Horizon}\label{sec:review}
\subsection{Geometry}
The near-horizon-geometry can be written as \cite{sss}
\bea
d s_{10}^2&=& d s_5^2+l^2\sum_{i=1}^3 \bigg{[}(d\mu_i)^2+\mu_i^2
  (d\xi_i+{2\over l\sqrt{3}}A)^2\bigg{]}\,,\\
F^{(5)}&=& (1+*_{(10)})\bigg{[}-{4\over l}{\rm vol}_5+{l^2\over
      \sqrt{3}}\sum_{i=1}^3 d(\mu_i)^2\wedge d\xi_i\wedge
    *_{(5)}F^{(2)}\bigg{]}\,,
\eea
where $\mu_1=\sin\alpha$, $\mu_2=\cos\alpha \sin\beta$ and
$\mu_3=\cos\alpha \cos \beta$ with $0\leq \alpha,\beta \leq \pi/2$,
$0\leq \xi_i\leq 2\pi$ and together they parametrise an S$^5$. 
Here in Poincar\'e coordinates for the AdS$_2$ part
\bea\label{eq:metric_poincare}
ds_5^2&=&-a^2 r^2 dt^2+b^2 \frac{dr^2}{r^2} +\frac{\omega^2}{ 4}\Bigl((\sigma_1^L)^2+(\sigma_2^L)^2\Bigr)+\frac{\omega^2}{4a^2
  b^2}\left(\sigma_3^L+{6 a^2 b^2\over l\omega}r dt\right)^2\,,
\eea
where $a^2=\frac{4 \lambda^2}{ \omega^2 l^2 \left(1+\frac{3\omega^2}{4
    l^2}\right)}$, $b^2={\omega^2 l^2\over 4 \lambda^2}$ and
  $\lambda=\sqrt{l^2+3\omega^2}$. The gauge potential is given by
\be
A=\frac{\sqrt{3}}{ 2} \left(\frac{2r}{\omega} dt+\frac{\omega^2}{4l}\sigma_3^L\right)\,.
\ee
The right-invariant one-forms on SU(2) are
\bea
\sigma_1^L&=&\sin\phi d\theta-\sin\theta \cos\phi d\psi\,,\\
\sigma_2^L &=& \cos\phi d\theta+\sin\theta \sin\phi d\psi\,,\\
\sigma_3^L&=& d\phi+\cos\theta d\psi\,.
\eea
The range of the angles are $0\leq \theta \leq \pi$, $0 \leq \psi
\leq 2\pi$ and $0\leq \phi \leq 4\pi$.
The 10-d Killing spinor is given by
\bea
\epsilon=\exp\left[-\frac{i}{2}(\xi_1+\xi_2+\xi_3)\right]\exp\left[-\frac{2i \lambda rt}{l\omega^2}
\Gamma_{49}(1+\Gamma_{09})\right] \times\nonumber\\ 
 \exp \left[\left(\frac{3 \omega}{4
  \lambda}\Gamma_{49}(1+\Gamma_{09})-\frac{1}{ 2}\Gamma_{09}\right)\ln r\right]
\epsilon_0\,,
\eea
where $\epsilon_0$ is a 32 component constant spinor satisfying
$\Gamma_{11}\epsilon_0=-\epsilon_0,
\Gamma_{0149}\epsilon_0=-i\epsilon_0,
\Gamma_{23}\epsilon_0=-i\epsilon_0, \Gamma_{57}\epsilon_0=-i\epsilon_0$. 
In terms of global coordinates for the AdS$_2$ part
\be\label{global5}
ds_5^2=-\left(1+{\rho^2\over b^2}\right)d\tau^2+{d\rho^2\over 1+{\rho^2\over
    b^2}}+{\omega^2\over 4}\Biggl((\sigma_1^L)^2+ (\sigma_2^L)^2\Biggr)+{\omega^2\over 4
  a^2 b^2}\left(\sigma_3^L-{6 ab \over \omega l}\rho d\tau\right)^2\,,
\ee and
\be \label{globalS5}
ds_{S^5}^2=l^2 \left(d\alpha^2+\cos^2\alpha d \beta^2+\sum_i \mu_i^2
(d\xi_i-{\omega^2 \over 4 l^2}\sigma_3 +{2 \over \omega l a b}\rho
d\tau)^2\right)\,,
\ee
with
\be
A=-{\sqrt{3}\over 2}\left(\frac{\omega^2}{4l}\sigma_3^L-\frac{2}{\omega a
  b}\rho d\tau\right)\,.
\ee
The global coordinate $\phi$ and Poincare $\phi$ are related by a
$\rho,\tau$ dependent transformation which leaves the period
invariant. In both coordinate systems, the geometry is that of $U(1)$ fibre bundle with coordinate $\phi$ over a
two-dimensional base sphere with coordinates $\theta,\psi$.
The Killing spinor is given by
\be
\epsilon=\exp\left[-{i\over 2}(\xi_1+\xi_2+\xi_3)\right]\exp\left[-{1\over
    2}\sinh^{-1} {\rho\over b} M\right]\exp\left[-{i\over 2}M
  \Gamma_{49}{\tau\over b}\right]\epsilon_0\,,
\ee
where $M={2b\over l}({3\over 2}\Gamma_{04}+{l\over \omega a b}
\Gamma_{09})$, $M^2=1$ and $\Gamma_{11}\epsilon_0=-\epsilon_0$,
$\Gamma_{0149}\epsilon_0=i\epsilon_0$,
$\Gamma_{23}\epsilon_0=i\epsilon_0$ and
$\Gamma_{57}\epsilon_0=-i\epsilon_0$\,.

In both coordinate systems there are four independent supersymmetries
that the geometry preserves which is twice the number that the full
black hole sees.


\subsection{Near-Horizon Probes}
In \cite{sss}, we investigated D3-brane probes  without world-volume fluxes in the near
horizon geometry. In the conventions of \cite{sss} there exist giant-like
anti-branes and dual giant-like branes in Poincar\'e coordinates, which preserve orthogonal
supersymmetries. In global coordinates there exist
$\BHG$ and $\tBHG$ solutions preserving the {\it same}
supersymmetries. Let us denote the world-volume coordinates by
$\sigma_0,\sigma_1,\sigma_2,\sigma_3$. The $\BHG$s have $\sigma_1=\beta$, $\sigma_2=\xi_2$ and
$\sigma_3=\xi_3$ while $\tBHG$s have
$\sigma_1=\theta$, $\sigma_2=\phi$, $\sigma_3=\psi$. In what follows,
we only review brane solutions for the case of global coordinates,
rather than anti-brane solutions.


\subsubsection*{Poincar\'e $\BHG$ and $\tBHG$ }
The Poincar\'e $\BHG$s and $\tBHG$s have $\sigma_0=t$. All the embedding coordinates are constant and
hence $H=0$. They preserve the supersymmetries obeying $\Gamma_{09}\epsilon_0=\epsilon_0$ and
hence are $\frac{1}{2}$-BPS with respect to the enhanced near-horizon
supersymmetries. In both cases the preserved
supersymmetry is the same as that of the full black hole. After integrating
over the world-volume spatial coordinates, the
expression for the non-zero conjugate momenta for $\BHG$ are 
\be
P_\phi=T_3 2\pi^2 l^2 \omega^2 \cos^2\alpha\,, \quad P_\psi=P_\phi
\cos\theta\,,\quad P_{\xi_1}=T_3 2\pi^2 l^4 \cos^2 \alpha.
\ee
Here $T_3$ is the D3-brane tension which we dropped in our earlier
paper. 

For the $\tBHG$ states the non-zero momenta are
\be
P_{\xi_i}= T_3 \pi^2 (\omega^2 l^2+(2 l^2+\omega^2) \omega^2)\mu_i^2\,,
\ee
The second term on the RHS arises from the WZ term and can be
gauged away. Another way of seeing this is to introduce a fictitious
parameter in front of it and note that the solutions are invariant
under a scaling of this parameter.
\subsubsection*{Global $\BHG$ and $\tBHG$}
Here $\sigma_0=\tau$ and for the $\BHG$ brane solutions, $\dot \phi=-{2l\over \omega
  \lambda}$ while $\dot\xi_1=-{2\omega \over l\lambda}$. Supersymmetry
dictates $\rho=0$, \textit{i.e.} the branes sit at the `centre' of the global $AdS_2$. The BPS
condition reads
\be
H_G=\frac{2l}{\omega \lambda} |\Pi_\phi|+\frac{2\omega}{l\lambda}
|\Pi_{\xi_1}|\,,
\ee
which is a function of $\alpha$. Here{\footnote{In the case of a point particle
    coupled to a gauge field the Hamiltonian is given by $(p-A)^2/2m$
    where $p=\partial L/\partial \dot x$. It is the combination of
    $(p-A)$ that ensures gauge invariance.}} and in what follows we have
defined $\Pi_x=P_x-A_x$ where $A_x$ is obtained from the WZ term by
writing it as $\dot x A_x$. Later, we will denote the energy density by
${\cal H}$ and the momentum densities corresponding to $\Pi_x$'s by
${\cal P}_x$, so that
\be
\Pi_x = \int_{\text{D3}} {\cal P}_x\,d\sigma_1d\sigma_2 d\sigma_3\,.
\ee
We find the non-zero momenta
\be
P_\phi=-T_3 \frac{2\pi^2}{3}l^4 \cos^2\alpha\,,\quad P_\psi=P_\phi\cos\theta\,,\quad P_{\xi_1}=-T_3
2\pi^2 l^4 \cos^2\alpha\,,
\ee
with $T_3$ being the D3 brane tension. 
In the case of the $\tBHG$s, $\dot \xi_i=-{2\omega\over l\lambda}$ and the
BPS condition reads
\be \label{dgH}
H_{DG}=\frac{2\omega}{ l
  \lambda}\left(|\Pi_{\xi_1}|+|\Pi_{\xi_2}|+|\Pi_{\xi_3}|\right)\,.
\ee
In this case $H_{DG}$ is a constant and the non-zero
momenta are
\be
P_{\xi_i}=-T_3 \pi^2 \omega^2 \left[{4\omega^2\over a^2
  b^2}-(\omega^2+2l^2)\right]\mu_i^2\,.
\ee
The second piece proportional to $(\omega^2+2l^2)$ comes from the
four-form potential and does not appear in the $\Pi$'s. Furthermore,
supersymmetry analysis dictates that there exists a gauge choice where
this term can be gauged away  and hence
$\Pi_{\xi_i}=P_{\xi_i}$ in this case.
In both cases, the preserved supersymmetry satisfies
$\epsilon_0^+=\Gamma_{49} \epsilon_0^-$ for branes with
$M\epsilon_0^\pm=\pm \epsilon_0^\pm$. 
The conserved spinor can be simplified to 
\be
\epsilon=e^{\mp{i\tau\over 2 b}}(1\mp\Gamma_{49})\epsilon_0^+\,,
\ee
where the upper sign is for branes and the lower sign for anti-branes.
The bilinear of this spinor leads to the BPS condition
\be\label{BPS}
H=\frac{2\omega}{ l\lambda}(|\Pi_{\xi_1}|+|\Pi_{\xi_2}|+|\Pi_{\xi_3}|)+{2l\over \omega
  \lambda} |\Pi_\phi|\,, 
\ee
where we identify $H=\partial_\tau$, $\Pi_{\xi_i}=\partial_{\xi_i}$ and
$\Pi_{\phi}=\partial_\phi$. 
\vskip.5em
{\it As is now clear, none of the sets of
branes without fluxes has all  four quantum numbers non-zero.}
\vskip.5em
The missing quantum number may be realised by electromagnetic or
mechanical waves. The former involves turning on world-volume fluxes
and latter deformations of the induced metric.
If the missing quantum number is to be provided by waves, then
(\ref{BPS}) predicts that for $\tBHG$ there should be a wave along
$\phi$ direction with velocity ${2 l\over \omega \lambda}$ while for
$\BHG$ there should a wave along $\xi_2+\xi_3$ direction with velocity
${2\omega \over l\lambda}$. We will see that this is precisely the
case.

\section{Counting Giants}

{\footnote{We thank N. Suryanarayana for collaboration in this section}}The promotion of the BPS condition (\ref{BPS}) to a quantum
condition suggests that the resulting quantum state may contain
both giant and dual giant parts. If there is a duality between the
two, which has yet to be established in the black hole context, then
it should be possible to describe the quantum states using dual giants
or giants alone. In this section we quantise the $\tBHG$ space of
solutions in global coordinates described above and compare the result
to the macroscopic entropy formula (\ref{entropy}). If we counted the
Poincar\'e $\tBHG$s we would get a divergence since all values of $r$
give the same energy.

The microstates of the black hole are conjectured to be a collection
of giant and/or dual giant gravitons. These branes correspond to
D3-dipoles and carry no net charge but they will still locally excite
the five-form field.  Hence when integrated over a small five-dimensional
surface which encloses a portion of the wrapped brane, the result will
be proportional to the number of D3-branes enclosed \cite{mt}.  With this picture in mind, we will integrate
components of $F$ over various spatial coordinates and use 
\be 
\int F = 16 \pi G_{10} T_3 \,n\,, 
\ee
with $n \in\mathbb{Z}$ in order to determine quantisation conditions. Using 
\be 
G_{10}={\pi^4 l^8\over 2
N^2}\,,\quad T_3={N\over 2\pi^2 l^4}\,, 
\ee
we have 
\be 
16 \pi G_{10} T_3 = {4\pi^3 l^4\over N}\,.  
\ee
Here $N$ is an integer obtained
after integrating $F_{\alpha \beta \xi_1 \xi_2 \xi_3}=4 l^4
\cos^3\alpha\sin\alpha \sin\beta\cos\beta$. Integrating 
\be
F_{\theta\phi\psi\alpha\xi_1}={q\over
16}\sin\theta\sin\alpha\cos\alpha 
\ee
with $q=-2 \omega^2(\omega^2+2
l^2)$ which is proportional to the electric charge gives
\be 
{N \over 2 l^4} \omega^2 (\omega^2+2 l^2)=n_1={N |q|\over 4 l^4}\,, 
\ee
while integrating 
\be 
F_{\phi\alpha\beta\xi_1\xi_2} = - \omega^2 l^2
\cos^3\alpha\sin\alpha\sin\beta\cos\beta\,, 
\ee
gives 
\be
{N\omega^2\over 2 l^2}=n_2\,.  
\ee
These together imply that
${N\omega^4\over 2 l^4}={2 n_2^2\over N}$ is also an integer. Note
that $n_1$ and $n_2$ are not independent but satisfy
$n_1=2n_2+2n_2^2/N$. In terms of $n_2$, the entropy can be rewritten
as 
\be 
S_{BH}=\pi\left({N\omega^2\over l^2}\right)^{3/2}\sqrt{N+{3 N \omega^2\over
4l^2}}=2\pi\sqrt{2 n_2^3\left(N+{3\over 2} n_2\right)}\,. \label{macro} 
\ee 
Here we have used $V_5=\pi^3 l^5$ and $G_{10}=V_5 G_5$.  We want to compare
 this entropy to a microscopic state counting using the microstates
 described in section \ref{sec:review}.
The gauge-invariant Hamiltonian for a single dual giant is given by
(\ref{dgH}). Furthermore, by solving the $\kappa$-symmetry constraint
as in \cite{sss,ms,masp,bm}, one can show that supersymmetry dictates the
following constraints
\be
\rho=0\,,\quad P_\rho=0\,,\quad\Pi_\alpha=0\,,\quad \Pi_\beta=0\,,\quad
\Pi_{\xi_i}-c\mu_i^2=0\,, 
\ee
where
\begin{equation} \label{eq:muconstraint}
\mu_1^2+\mu_2^2+\mu_3^2=1\,,
\end{equation}
 which can be treated as an additional  constraint. Here
\be
c=-{V_3\over 8\pi^2}{N \omega^4 \over l^4}\left(1+{3\omega^2\over 4
  l^2}\right)=-{V_3\over 8\pi^2}{4
  n_2^2\over N}\left(1+{3n_2\over 2 N}\right)\,.
\ee
Here $V_3$ is the volume factor obtained after integrating over the
spatial world-volume coordinates. The integration over the full range
gives $16\pi^2$. We will leave it undetermined for now.
Following Dirac's procedure for 2nd class constraints,
one can simply drop $\rho,P_\rho$ from the phase space. After
quantisation, the remaining constraints can be thought to be imposed on the
Hilbert space satisfying the gauge-invariant bracket $[\Pi_{a},x^a]=-i$. Demanding this canonical
commutation relation gives us
\be
[c \mu_i^2,\xi_j]=-i\delta_{ij}\,.
\ee
Defining the classical variables $\zeta_i=\sqrt{|c|}\mu_i e^{i\xi_i}$ and
promoting them to quantum operators gives us the oscillator
brackets{\footnote{The role of creation and annihilation operators
    gets interchanged when considering anti-branes.}}
\be
[\zeta_i,\zeta_j^\dagger]=\delta_{ij}\,.
\ee
This leads to writing the quantum Hamiltonian as 
\be
H = {2\omega\over l\lambda} (\zeta_i^\dagger
\zeta_i)={2\omega\over 
l\lambda}(N_1+N_2+N_3)\,.
\ee
Now imposing the restriction $\mu_1^2+\mu_2^2+\mu_3^2=1$ we see that the quantum states created by these oscillators are
\begin{equation}
|N_1\,,N_2\,,N_3\rangle=\prod_{i=1}^3\frac{(\zeta_i^\dagger)^{N_i}}{\sqrt{N_i !}} |\,{\rm vac}\rangle
\end{equation}
with occupation numbers satisfying
\be
N_1+N_2+N_3=|c|\,.
\ee
Thus we have constructed the Hilbert space of a constrained three-dimensional harmonic oscillator.

Instead of directly imposing the quantum commutator brackets we can also proceed by applying
Dirac's procedure to deal with second-class constraints \cite{Das:2000fu,Mandal:2005wv,ms,masp}. Imposing (\ref{eq:muconstraint}) on the classical phase space implies the relation
$$\Pi_{\xi_1}+\Pi_{\xi_2}+\Pi_{\xi_3}=c\,,$$
 which is a first-class constraint. Thus we can take
\begin{equation}
 \Pi_\alpha=0\,,\quad\Pi_\beta=0\,,\quad\Pi_{\xi_2}=c\mu_2^2\,,\quad\Pi_{\xi_3}=c\mu_3^2
\end{equation}
as a system of second-class constraints. We define the Poisson brackets as $\{f,g\}_{PB}={\partial f\over \partial \Pi}{\partial g\over \partial x}-{\partial f\over \partial x}{\partial g\over \partial \Pi}$, which is the classical equivalent to the quantum condition  $[\Pi_{a},x^a]=-i$. This procedure can be justified by realizing that there exists a gauge for the four form $C^{(4)}$ and thus for the effective gauge potential $A$, in which the term giving rise to the $q$-piece in the momentum constraint drops out. Following this procedure,  we get the following commutator
brackets
\be
[c\mu_p^2,\xi_q]=-i \delta_{pq}\,,\quad p,q=2,3\,.
\ee
With these we can define two oscillators $\zeta_2=\sqrt{|c|}\mu_2 e^{i \xi_2},\zeta_3=\sqrt{|c|}\mu_3 e^{i\xi_3}$ which satisfy the algebra
of two commuting simple harmonic oscillators. This then yields
\be
|\Pi_1|=|c|-N_2-N_3\,,
\ee
as before. Thus we again end up with the Hilbert space of a constrained three-dimensional harmonic oscillator, whose state counting is a three-coloured partitioning problem.

Integrating over a five-dimensional surface transverse to the dual-giant
world volume will give us the total number of dual giants allowed in
the geometry. The transverse coordinates are $\alpha,\beta,\xi_i$ and
this leads to the maximum number of dual giants to be $N$.  When we
consider $M$ multiple dual giant probes, we need to satisfy \cite{ms,masp,bm}
\be
\sum_i^{M} \left(N_1^{(i)}+N_2^{(i)}+N_3^{(i)}\right)=M |c|\,.
\ee
In terms of the integer $n_2$ and $N$ the right-hand side can be rewritten as
\be
{V_3\over 8\pi^2}{4 M n_2^2 \over N^2}\left(N+{3\over 2} n_2\right)\,.
\ee
We need the three-coloured partition of this in the limit $N\gg M\gg 1$ which
will give the entropy
\be
S_{probes}^{\tBHG}=2\pi\sqrt{{V_3\over 8\pi^2}{2 M n_2^2 \over N^2} \left(N+{3\over 2} n_2\right)}\,.\label{stat}
\ee
Note that for this argument to make sense we need to make sure that
the integer we are partitioning is much less than $M$ as this is the
upper limit on the sum. 
This leads to the condition $\omega \ll l$.
When $M=N$ and with $V_3=16\pi^2$,  we can associate
this factor with the Landau degeneracy of $\BHG$. For giants 
\begin{align}
\Pi_\phi&=P_\phi-A_\phi=P_\phi+T_3 {\pi^2\over 2} \omega^2 l^2
\cos^4\alpha\,,\nonumber \\
\Pi_{\xi_1}&=P_{\xi_1}-A_{\xi_1}=P_{\xi_1}-2 T_3
\pi^2 l^4 \cos^4\alpha\,.
\end{align}
Thus the maximum integral quantum number associated with the state annihilated
by  $P_\phi$ is $n_2/2$
and that with $P_{\xi_1}$ is $N$. There is an additional factor of 2
corresponding to the additional giant solutions \cite{sss} found at
$\theta=0,\pi$ which carry the same quantum numbers. In total we have
a degeneracy factor of $Nn_2$. It
remains to be seen if this is merely a coincidence. One possibility is
that this degeneracy is to be perceived as the ground state degeneracy
for each dual giant and hence $3 N n_2$ colours rather than 3. Putting
together all the ingredients above, we get $S_{probes}=2 S_{BH}$. 

Eventually we would like a more rigorous justification for this missing
degeneracy we observed above. The next obvious question to ask is: What happens when one
switches on world-volume electromagnetic flux? Since this is known to
provide angular momentum, it is natural to suspect that the missing
quantum number, in this case associated with $\Pi_\phi$ may arise
from the electromagnetic field. Let us suggest the following way of
counting motivated by \cite{kmmr} which leads to the same relation
between $S_{probes}$ and $S_{\text{BH}}$ as in this section. 


\subsection*{Adding a fourth quantum number}
When we have $\Pi_\phi$ turned on, either by electromagnetic waves (as
shown in section \ref{sec:wvsusy}) or otherwise, the BPS relation suggests
\be \label{constr1}
\Pi_\phi+{\omega^2 \over l^2} (\Pi_{\xi_1}+\Pi_{\xi_2}+\Pi_{\xi_3})=P\,,
\ee
where $P$ denotes the total momentum. Meanwhile, we have from the probe analysis
\be
\Pi_{\xi_1}+\Pi_{\xi_2}+\Pi_{\xi_3}=n\, .
\ee
This partitioning of $n$ into three integer-valued momenta can be accomplished in $n^2/2$ ways when $n$ is large. This can
be achieved by taking $\omega\ll l$, but $N\gg 1$. As we
will show in later sections $\Pi_\phi$ can be constructed out of two
integers and hence keeping $P$ fixed can be realized in
$P-{\omega^2\over l^2}n$ ways.
Thus the total number of ways of satisfying the above conditions is given by
\be
\frac{n^2}{2}\left(P-\frac{\omega^2}{l^2} n\right) \,,
\ee
which is maximised w.r.t $n$ for $\Pi_\phi={\omega^2 \over 2
  l^2}n$ which can be small compared with $n$ and hence can be thought
of as arising from small fluctuations. Now we anticipate that this momentum is going to be carried
by open strings which are $MN$ in number since there are $M$ dual
giant probes and $N$ dual giants making the black hole. The bosonic moduli corresponding to
$\alpha,\beta$ and fermionic moduli corresponding to the 2 preserved
supersymmetries will contribute a factor of 3. The microscopic entropy
arising from the partitioning of $\Pi_\phi$ is given by

\be\label{eq:Sprobes}
S_{probes}=2\pi \sqrt{{3 \Pi_\phi MN\over 6}}=4 \pi {M\over N} \sqrt{2
  n_2^3\left(N+{3\over 2} n_2\right)}=2 {M\over N}S_{BH}\,,
\ee 
when{\footnote{There may be an overall $O(1)$ factor
having to do with the subtlety in counting independent open
string states stretched between giants (see \cite{
Balasubramanian:2004nb}) present.}} $n=M|c|$. Let us now explain why this relation is expected. 


\subsection*{What does $S_{probes}$ count?}
Let us observe that $S_{BH}$ scales in terms of the number of dual
giants $N$ like $ S_{BH} = f\left( \omega,l  \right)\,N^2$.
 In the analysis leading to (\ref{eq:Sprobes}) we computed the entropy associated with
inserting $M$ probe branes into the near-horizon geometry of the black
hole. When we insert $M$ probes in the black hole geometry, these
will form a new bound state with a higher entropy proportional to
$(M+N)^2$. The open string degrees of freedom associated with this new
bound state are $MN$ in number. The $M^2$ and $N^2$ open strings
ending on the same type of branes take into account the degrees of
freedom associated with separating the objects. We are associating the degeneracy of the
probes with the number of ways that the $MN$ open strings can carry
$\Pi_\phi$. 
Then our computation should correspond to the difference in the
entropy of the new bound state made of $M+N$ branes and the entropy
when the probe and the black hole are far apart. This is given by 
\begin{equation}
S_{probes} =  f(\omega,l) \left[(N+M)^2 - N^2 - M^2  \right]=2 NM
f(\omega,l)=2 {M\over N} S_{BH} \,.
\end{equation}
Taking
$N=M$, we arrive at the conclusion that $S_{probes}=2S_{BH}$,
the result that emerged from two independent computations
above. We
emphasise that our identification of dual giants gravitons and black hole
microstates is conjectural, but we take the above results as
encouraging evidence for such a connection. Eventually, it will be
important to understand why dual giants, which are objects expanding
in $AdS$ are a valid microscopic description of a black hole.

We must also remind the reader that we have not demonstrated the counting by quantisation of the
phase space of the BPS waves directly which we will leave as an open
problem. However, the existence of waves carrying the right velocity
which preserve the same supersymmetry as the non-fluxed solution makes
it very plausible that the above argument is at least on the right track.

To provide more evidence, we now need to demonstrate the
existence of BPS modes carrying $\Pi_\phi$ which we turn to in the next
section.



\section{Supersymmetry and World Volume Fluxes}\label{sec:wvsusy}
In this section we want to investigate the possibility of preserving some fraction of supersymmetry
for D3-branes with non-trivial world-volume gauge field
configurations\footnote{For earlier work on world-volume fields in the
context of giant gravitons see \cite{Sadri:2003mx}.}. These are governed by an action of the form
\begin{equation}\label{eq:actionD3}
L = -T_3\int \sqrt{-\det(h+F)} \,\text{d}^4\sigma\mp T_3\int C^{(4)}\,,
\end{equation}
where in accordance with \cite{sss}, the upper sign stands for a brane and the lower sign for an anti-brane and $C^{(4)}$ is the pull back of the space-time four form potential.
We shall investigate the question of supersymmetry from the point of
view of world-volume $\kappa$-symmetry transformations. In the
presence of world-volume flux, the supersymmetry condition for a
D3-brane is \cite{3brane,Bergshoeff:1996tu}
\be\label{eq:kappasymmetry}
\Gamma\epsilon=\epsilon\,,
\ee
with the general $\kappa$-symmetry projector
\be\label{eq:fluxGamma}
\Gamma={\epsilon^{ijkl}\over \sqrt{-{\rm det}(h+F)}}\left({1\over
    4!}\gamma_{ijkl}I-{1\over 4}F_{ij}\gamma_{kl} J+{1\over 8}F_{ij}F_{kl}I\right)\,,
\ee
where
\bea
I\epsilon&=&-i\epsilon\,,\\
J\epsilon&=&i\epsilon^*\,.
\eea
For an anti-brane the right hand side of (\ref{eq:kappasymmetry}) has the opposite sign.
Note that this simplifies to the condition (6.2) of \cite{sss} in the absence of world-volume fluxes, as required.
Since we want to preserve the same supersymmetries as in the $F=0$ case in \cite{sss}, we must demand that
\be \label{eq:kappaconditions}
\epsilon^{ijkl}F_{ij}\gamma_{kl}\epsilon^*=0,\qquad \epsilon^{ijkl}F_{ij}F_{kl}=0\,.
\ee
Define the world-volume field strength tensor as 
\begin{equation}
F=\left(\begin{array}{cccc}
0 & E_1&E_2&E_3\\
-E_1 & 0 & B_3 & -B_2 \\
-E_2 & -B_3 & 0 & B_1\\
-E_3 & B_2 & -B_1 & 0
  \end{array}
\right)\,.
\end{equation}
Since we are in four space-time dimensions we can split $F$ into electric and magnetic fields. Then the second of the conditions (\ref{eq:kappaconditions}),\textit{ i.e.} $F\wedge F=0$, implies that $\mathbf{E}$ and $\mathbf{B}$ are orthogonal to each other. The first condition above implies
\begin{equation}\label{eq:fluxsusycondition}
(E_1 \gamma_{23}-E_2 \gamma_{13}+E_3 \gamma_{12}+B_3 \gamma_{03}+B_2
\gamma_{02}+B_1\gamma_{01})\epsilon^*=0\,.
\end{equation}
After solving the supersymmetry constraint, it is still necessary to
check the equations of motion. The embedding coordinates' equations of
motion follow from varying the action (\ref{eq:actionD3}). We choose
to work in static gauge, aligning the four world-volume coordinates
with certain space-time coordinates. Which set of space-time
coordinates we choose will vary from case to case. The gauge field
equations of motion are compactly given by the expression \cite{Aganagic:1996nn}
\begin{equation}\label{eq:gaugefield}
\partial_i \left(\sqrt{-\det(h+F)}\left\{ (h+F)^{-1} - (h-F)^{-1} \right\}^{ij} \right)=0\,.
\end{equation}
Finally, the Bianchi identities of the world-volume gauge fields, $d F=0$ must also be satisfied.

As a warmup to the near-horizon geometry, but also because the result
is interesting in its own right, we shall now analyse dual giant
gravitons in AdS$_5\times$S$^5$ with world-volume fluxes. In
\cite{Kim:2005mw}, giant-graviton configurations in
AdS$_5\times$S$^5$ were constructed following the method of Mikhailov \cite{mikhailov}. There is was found that it is possible to excite electric and magnetic fields on the brane without breaking any further supersymmteries. The gauge fields obey wave equations and contribute a momentum to the BPS relation via their Poynting vector. We realise this scenario on dual giant gravitons and find that turning on fluxes on dual $\tfrac{1}{8}$-BPS giant gravitons breaks the supersymmetry further to $\tfrac{1}{16}$, at least for the type of configuration we study.
\subsection{Fluxes on $\frac{1}{8}$ BPS Dual Giants in AdS$_5\times$ S$^5$}
We will closely follow \cite{ms} and the reader is referred to it for
more details. The AdS$_5\times $S$^5$ metric is
\begin{equation}\label{eq:AdS5S5metric}
ds^2=-Vdt^2+\frac{1}{V}dr^2+\sum_{i=1}^3\frac{r^2}{4}(\sigma_i^L)^2+l^2(d\alpha^2+\cos\alpha^2 d\beta^2+\sum_{i=1}^3 \mu_i^2 d\xi_i^2)\,,
\end{equation}
where $V=1+{r^2\over l^2}$, $\mu_1=\sin\alpha$, and  $\left\{\mu_i\right\}$ and $\left\{\sigma_i^L\right\}$ have the same meaning as in section \ref{sec:review}.
After a coordinate transformation we can write the 3-sphere metric in the alternative form
\be\label{S32}
r^2 (d\theta^2+\cos\theta^2 d\phi_1^2+\sin\theta^2 d\phi_2^2)\,.
\ee
The world-volume coordinates are labelled by $\sigma_i$ with $i=0,1,2,3$. We
choose static gauge such that $t=\sigma_0, \theta=\sigma_1,\phi=\sigma_2,\psi=\sigma_3$.
The world-volume gamma matrices for AdS$_5$ dual giants in the coordinates of (\ref{eq:AdS5S5metric}) are
\begin{align}\label{pull1}
\gamma_0=V^{1/2} \Gamma_0+\sum \mu_i \Gamma_{6+i}\,,\quad \gamma_{1}= \frac{r}{2} (\sin\phi \Gamma_2+\cos\phi \Gamma_3)\,,\nonumber\\
\gamma_{2}= \frac{r}{4}\Gamma_4\,, \quad\gamma_{3}=
\cos\theta\gamma_2-\frac{r}{4}\sin\theta(\sin\phi \Gamma_3-\cos\phi\Gamma_2)\,.
\end{align}
%
In terms of the coordinates of  (\ref{S32}), \textit{i.e. } $\{t,r,\theta,\phi_1, \phi_2  \}$ these gamma matrices are
\begin{align}\label{pull2}
\gamma_0=V^{1/2} \Gamma_0+\sum \mu_i \Gamma_{6+i}\,,\quad\gamma_{1}= r \Gamma_2\,,\nonumber \\
\gamma_{2}= r\cos\sigma_1 \Gamma_3\,,\quad \gamma_{3}= r\sin\sigma_1\Gamma_4\,.
\end{align}
%
%
%
In these coordinates the induced metric is diagonal. $\frac{1}{8}$-BPS dual giants without gauge fields satisfy
\be
\gamma_0\gamma_1\gamma_2\gamma_3\epsilon=-i\sqrt{-\det h}\,\epsilon\,,
\ee
where $h$ is the induced metric on the world-volume of the brane.
The spinor $\epsilon$ is subject to the projection conditions
\begin{equation}
\Gamma_{09}\epsilon=\epsilon\,,\quad
\Gamma_{68}\epsilon=i\epsilon\,,\quad \Gamma_{57}=i\epsilon\,.
\end{equation}
We now wish to preserve a fraction of supersymmetry with non-trivial gauge fields. In order to solve (\ref{eq:kappasymmetry}) with $\Gamma$ given by (\ref{eq:fluxGamma}), we
need to satisfy
\be
\epsilon^{ijkl}F_{ij}\gamma_{kl}\epsilon^*=0\,.
\ee
In terms of (\ref{pull2}) this condition becomes
\begin{equation}
\left[\left(E_1-i\frac{\sqrt{-h}}{ h_{22}h_{33}}B_1\right)\gamma_{23}-\left(E_2-i\frac{\sqrt{-h}}{h_{11}h_{33}}B_2\right)\gamma_{13}+\left(E_3-i\frac{\sqrt{-h}}{h_{11}h_{22}}B_3\right)\gamma_{12}\right]\epsilon^*=0\,.
\end{equation}
By equating real and imaginary part of this equation to zero
individually it follows that without imposing any further projection
conditions, we need to set ${\bf E}={\bf B}=0$. However, if we impose
the additional projection $\Gamma_{23}\epsilon=i\epsilon$ in the basis
(\ref{pull1}), then gauge fields obeying
\begin{equation}\label{eq:AdS5S5constraints}
E_2=0\,, \quad B_3\cos\theta=-B_2\,,\quad E_1=-{2\over l}B_3\,,\quad
E_3={2\over l}B_1\,.
\end{equation}
solve the $\kappa-$symmetry condition including world-volume fluxes.
Because of the extra projection condition, turning on the gauge field leads to breaking more supersymmetries. In the specific case above it leads to $\frac{1}{16}$-BPS states. As a consequence of supersymmetry, this configuration has 
$ {\bf E}\cdot {\bf B}=0\,.$
 The equations of motion for the embedding coordinates are solved by $\dot\xi_i={1\over l}, \dot
\alpha=\dot\beta=\dot r=0$, if the fields satisfy the equation
\begin{equation}
\partial_0 B_i-{2\over l}\partial_2 B_i=0\,, i=1,3\,.
\end{equation}
Hence we have waves moving with phase velocity $2/l$ in the $\phi$ direction.
Finally, the gauge field equations of motion with the Bianchi
identities give
\begin{eqnarray} \label{ads5eom}
\sin\theta\partial_\theta (E_1 \sin\theta)-\cos\theta \partial_{\phi} E_3+\partial_\psi E_3&=&0\,,\nonumber\\
\partial_\theta E_3-\partial_\psi E_1+\partial_{\phi} E_1 \cos\theta&=&0\,.
\end{eqnarray}
As we will show in the next section these equations can be expressed
 compactly in terms of $\mathbb{C}P^1$ coordinates when the 3-sphere metric is
explicitly written as a Hopf fibration. In this picture the waves propagate along the fibre.
With the constraints (\ref{eq:AdS5S5constraints}), we have the
dramatic simplification
\begin{equation}
\sqrt{-\det h}=\sqrt{-\det(h+F)}\,.
\end{equation}
 These are analogous to the waves on $\frac{1}{8}$-BPS giant gravitons analysed by \cite{Kim:2005mw} with the difference that there the inclusion of waves did not break any further supersymmetries. 
The gauge field contributes canonical momenta
\begin{equation}
{\cal P}_{E_1} = \frac{\partial {\cal L}}{\partial E_1}= T_3 B_3\sin\theta\,,\quad {\cal P}_{E_2} = \frac{\partial {\cal L}}{\partial E_2}= T_3 B_1\cot\theta \,,\quad {\cal P}_{E_3} = \frac{\partial {\cal L}}{\partial E_3}= -T_3 B_1\csc\theta\,.
\end{equation}
The Hamiltonian density is given by
\begin{equation}
{\cal H}=\frac{1}{l}\left[2|{\cal P}_\phi| +\sum_{i=1}^3|{\cal P}_{\xi_i}|\right]\,,
\end{equation}
where
\begin{equation}
{\cal P}_\phi = B_1{\cal P}_{E_3} - B_3{\cal P}_{E_1}\,.
\end{equation}
and
\begin{equation}
{\cal P}_{\xi_i}= -T_3 \left[\frac{l^2}{r^2\sin\theta}(B_1^2+B_3^2\sin^2\theta)\mu_i^2+r^2 l^2\mu_i^2\sin\theta\right]\,.
\end{equation}
The angular momentum of the gauge field has introduced a new quantum number in addition
to $(J_1,J_2, J_3)$ leading to the four-tuple  $(S_1,J_1,J_2,J_3)$. 
When counting the degeneracy of such states, one focuses on states of fixed energy. Since the Hamiltonian is $r$ dependent, there is a certain energy for each $r,P_\phi$ . The total number of ways of choosing $r,P_\phi$ to achieve this energy after quantisation corresponds to the degeneracy of these solutions. If $P_\phi=0$ then each value of $r$ corresponds to a different
energy and the degeneracy is unity. With $P_\phi\neq 0$ turned on, we get a larger
degeneracy since now different choices for $r,P_\phi$ can give the same energy. 
It would be interesting to carry out the quantisation of the new phase
space and count these objects. We will not attempt to do so in this
paper.
However, let us attempt to motivate how these asymptotic states could
be used to account for the microscopic entropy. Firstly, we have
\be
E l=2 P_\phi+P_{\xi_1}+P_{\xi_2}+P_{\xi_3}\,,
\ee
with
\be
P_{\xi_1}+P_{\xi_3}+P_{\xi_3}=n\,,
\ee
which can be realized in $n^2/2$ ways.
It is natural to identify $E$ with the mass of the black hole which is
known to be
\be
M={3\pi\omega^2\over 4 G_5}\left(1+{3\omega^2\over 2 l^2}+{2\omega^4\over 3
  l^4}\right)\,.
\ee
The total number of ways in which the above constraints can be
satisfied is $(E l-n)n^2/2$ ways which is maximised when $E l=(3/2)
n$. Comparing now the mass of the black hole with this, we have
\be
n={\omega^2 N^2 \over l^2}\left(1+{3\omega^2\over 2 l^2}+{2\omega^4\over 3
  l^4}\right)\,,
\ee
with $P_\phi=n/4$. Assuming now that this is carried by $N^2$ open
strings with a central charge of 3 arising from $\alpha,\beta$ and 2
supersymmetries, we have the microscopic entropy given by
\be
S_{micro}=2\pi\sqrt{{\omega^2 N^4\over 4 l^2}\left(1+{3\omega^2\over 2
    l^2}+{2\omega^4\over 3 l^4}\right)}\,,
\ee
which agrees with $S_{BH}$ to leading order when $\omega \ll l$ but
differs at higher orders. It will be nice to derive the analogous
formula by quantising the phase space of solutions rather than by this
indirect way.

\subsection{General Solution}
It will turn out that the differential equations obeyed by the $\BHG$ and $\tBHG$ configurations we are about to investigate can be transformed into an equivalent form both in Poincar\`e and global coordinates.
Before we analyse particular instances of $\BHG$ and $\tBHG$ configurations, we present here the general solution to these equations.
Let us introduce the complex variable
\begin{equation}
z=2 e^{i\psi} \tan {\theta\over 2}\,.
\end{equation}
Then
\begin{equation}
\partial_\theta={1\over \sqrt{z \bar z}}\left(1+{z \bar z\over 4}\right)(z
\partial_z +\bar z \partial_{\bar z})\,,
\end{equation}
and 
\begin{equation}
\partial_\psi=i z \partial_z-i \bar z \partial_{\bar z}\,.
\end{equation}
For later convenience let us briefly describe the geometry of these coordinates, in terms of which the metric on a squashed three sphere of radius $R$ reads
\begin{equation}
d \Omega_3^2 = \frac{R^2}{4} \left[  \frac{16dzd\bar z}{(4 + z \bar z)^2} + q^2 (d\phi + {\cal A}(z,\bar z))^2 \right]\,,
\end{equation}
where ${\cal A}$ is a one form that lives purely in the $\mathbb{C}P^1$ base, parametrised by $z$ and $\bar z$. We have
\begin{equation}
 {\cal A} = \frac{1}{2iV} \left( z^{-1}dz -  \bar z^{-1} d\bar z \right)\,.
 \end{equation}
Here
\begin{equation}
V={4+z \bar z\over 4-z \bar z}\,.
\end{equation}
The squashing parameter $q$ is unity for the round three sphere and is
determined for the solutions, together with the radius $R$,  in terms of the AdS length $l$ and rotation parameter $\omega$.
The equations we want to solve take the form (\ref{ads5eom}) and can be
compactly written as
\begin{equation}
2 V z \partial_z G=-i \partial_\phi G\,, \qquad 2 V \bar z \partial_{\bar z}
\bar G=i \partial_\phi \bar G\,, \label{eq:complex}
\end{equation}
where $G$ is a complex field, in the AdS$_5$ case of the previous
section,  $G=E_3+i \sin \theta
E_1$. Evidently one is the complex conjugate of the other. We now obtain the general solution of (\ref{eq:complex}).
We expand $G(z,\bar z;\phi)$ in eigenmodes of the $\partial_{\phi}$ operator, keeping in mind the $4\pi$ periodicity of $\phi$:
\begin{equation}
G(z,\bar z;\phi) = \sum_{k=-\infty}^\infty G_k(z,\bar z) e^{-\frac{ik}{2} \phi}\,.
\end{equation}
This leads to the equation
\begin{equation}
\partial_z \ln G_k(z,\bar z) = \frac{ik}{2} {\cal A}_z (z)\,
\end{equation}
with solution
\begin{equation}\label{eq:Gksolution}
G_k =  \bar g_k(\bar z) \exp\left[{\frac{ik}{2}\int {\cal A}_z dz} \right]\,.
\end{equation}
Here $\bar g_k(\bar z)$ is an arbitrary anti-holomorphic function, i.e. independent  of $ z$. Regularity at $\theta=0,\pi$ dictates that it take the form
\begin{equation}
\bar g_k(\bar z) =\sum_{n=-k/2}^{k/2}a_{k,n}\bar z ^n\,.
\end{equation}
Here $k$ is an integer, so that the allowed values for $n$ are integers and half-odd integers.
giving a degeneracy of $2k+1$ for each $k$. Thus for a given $\phi$
momentum $k$ we have a degeneracy of $2k+1$ in the sense that there
are $2k+1$ ``independent'' coefficients that determine $\bar G$.  The
electromagnetic fields on the $\BHG$s and $\tBHG$s may be quantised by treating them as small fluctuations around the zero-field vacuum in a fashion analogous to \cite{kl}. Upon quantisation the expansion coefficients $a_{k,n}$ and $a_{k,n}^*$ become creation and annihilation operators, from which we may construct two additional number operators that correspond to the excitations of the complex field $G$.

The integral in (\ref{eq:Gksolution})  may be done explicitly yielding
\begin{equation}
G_k(z,\bar z) = \bar g_k(\bar z)\left[  \frac{z\bar z}{(4 + z\bar z)^2} \right]^{\frac{k}{4}}\,.
\end{equation}

\subsection{Black Hole Giants with Fluxes}
We now turn to specific examples of compact D3-brane configurations
with non-trivial world-volume fluxes in the near-horizon geometry. The
analysis in the sections below applies to the case of a brane. 
\subsubsection{Global $\tBHG$}\label{sec:globaldualgiants}
Consider a D3-brane with world-volume coordinates $\mathbf{\sigma}=\{\tau,\theta,\phi,\psi\}$
in static gauge. Furthermore, we assume that the embedding coordinates
$X^m(\sigma_0)$ depend on time only and obey
\be
\dot\psi=0,\quad \dot\phi=0,\quad \dot\xi_i=-\frac{2\omega}{l\lambda}\,\quad \dot\alpha=0\,\quad\dot\rho=0
\ee
In the absence of flux, supersymmetry further dictates $\rho=0$, a
feature that carries to the fluxed solutions.
It can be shown (see appendix for details of the computation) that  (\ref{eq:fluxsusycondition}) leads to the condition that the fields satisfy
\begin{equation}
E_2=0\,,\quad B_3 \cos\theta=-B_2\,\quad E_1=-{2l\over \omega \lambda} B_3\,,\quad E_3={2l \over \omega
  \lambda} B_1\,.
\end{equation}
With these constraints there occurs a significant simplification of the on-shell DBI action. We find that
\begin{equation*}
\sqrt{-\det{(h+F)}} =\sqrt{-\det{h}}\,.
\end{equation*}
Furthermore, it is evident that the field configurations above satisfy $\mathbf{E}\cdot \mathbf{B}=0\,.$
From these relations it follows that these solutions preserve the \textit{same} supersymmetries as the un-fluxed case found in \cite{sss}. We now demonstrate that the above configurations are indeed solutions to the equations of motion subject to certain further equations that can be solved in general.
The equations of motion for the embedding coordinates can be shown to be satisfied if the two independent components (we choose to solve the constraints for $E_1$ and $E_3$) of the field strength satisfy
\be\label{coordeq}
\partial_0 E_i-{2l\over \omega \lambda} \partial_2 E_i=0\,, \quad
i=1,3\,.
\ee
From the supersymmetry constraints above it follows that the $B_i$ satisfy a set of analogous equations.
In addition to these we must also make sure that the gauge field on the brane obeys the Bianchi identities
\bea
\partial_2 E_1-{\omega\lambda\over 2 l}\partial_0 E_1 &=& 0\,,\qquad \partial_2 E_3-{\omega\lambda\over 2 l}\partial_0 E_3 =0\,,\nonumber\\
\partial_3 E_1-\partial_1 E_3 -{\omega\lambda\over 2 l}\partial_0 E_1\cos\theta &=& 0\,,\qquad \partial_3 E_1-\partial_1 E_3-\partial_2 E_1\cos\theta=0\,.\nonumber\\
\eea
and equations of motion (\ref{eq:gaugefield}). The first two are identical to the coordinate equations of motion. Combining the non-trivial information from the Bianchi identities with the gauge-field equations of motion leaves us with solving the system of partial differential equations
\begin{eqnarray}\label{eq:globaldualgiant}
\sin\theta\partial_1 (E_1 \sin\theta)-\cos\theta \partial_2
E_3+\partial_3 E_3&=&0\,,\nonumber \\
\partial_1 E_3-\partial_3 E_1+\partial_1 E_1 \cos\theta&=&0\,.
\end{eqnarray}
Note that these are precisely the same as (\ref{ads5eom}). The time
dependence is given by (\ref{coordeq}), so that $E_i (\tau,\phi,\psi)=
E_i(\sigma^+,\psi)$, where we have defined the light-cone variable
$\sigma^+ = \frac{2l}{\omega\lambda}\tau + \phi$. 
\vskip.5em
\textit{Thus, physically, these solutions correspond to waves
  travelling with a phase-velocity that is exactly in accordance with
  the general BPS relation (\ref{BPS}). } 
\vskip.5em
The gauge field gives rise to the conjugate momentum densities
\begin{align}
{\cal P}_{E_1} = \frac{\partial {\cal L}}{\partial E_1} = -T_3 B_3 \sin\theta\,,\quad{\cal P}_{E_2} = \frac{\partial {\cal L}}{\partial E_2} = -T_3 B_1 \cot\theta\,, \quad{\cal P}_{E_3} = \frac{\partial {\cal L}}{\partial E_3} =  T_3 B_1 \csc\theta\,.
\end{align}
The Hamiltonian density is
\begin{equation}\label{eq:globdual_hamiltonian}
{\cal H}=\frac{2l}{\omega \lambda}|{\cal P}_\phi|+\frac{2\omega}{l\lambda}\sum_{i=1}^3|{\cal P}_{\xi_i}|\,,
\end{equation}
where ${\cal P}_{\xi_i}$ denote unintegrated $\Pi_{\xi_i}$ with
\begin{equation}
{\cal P}_\phi=B_1 {\cal P}_{E_3}-B_3 {\cal P}_{E_1}
\end{equation}
and
 \begin{align}
 {\cal P}_{\xi_i} &= - T_3\left( \frac{\omega}{4l}\right)^2 \mu_i^2 \csc\theta \left[ 12\lambda^2 |G|^2 + \omega^2(4l^2 + 3\omega^2)\sin^2\theta \right]\,,
 \end{align}
We have defined the quantity 
\begin{equation}
G = E_3 + i\sin\theta E_1\,.
\end{equation}

Equation (\ref{eq:globdual_hamiltonian})  reproduces the BPS condition (\ref{BPS}) with all four charges. Notice that three of the charges are realized as `orbital' angular momenta of the classical brane motion, whereas one is realized in terms of angular momentum carried by the gauge field on the brane.
Rewriting (\ref{eq:globaldualgiant}) in terms of the new complex variables and taking linear combinations leads precisely to equations (\ref{eq:complex}), whose solutions were obtained above.

\subsubsection{Poincar\'e $\tBHG$}
We shall now work in the coordinate system (\ref{eq:metric_poincare}). Let us consider a $D3$-brane with world volume coordinates $\mathbf{\sigma}=\{
t, \theta, \phi,\psi\}$,
where $t$ is AdS$_2$ Poincar\'e time as defined in \cite{sss}. We
assume static gauge and in addition that the remaining embedding
coordinates are functions of $\sigma_0$ only. They satisfy
\be
\dot\xi_i=\dot\alpha=\dot\beta=0
\ee
From the analysis in \cite{sss} it follows that these satisfy $\gamma_0\epsilon^*=0$. Using this, we find that (\ref{eq:fluxsusycondition}) implies
\begin{equation}
E_i=0\,,\qquad B_2 + B_3 \cos\theta=0\,
\end{equation}
with $B_1$ unconstrained by supersymmetry. Here, the equations of motion and Bianchi identities reduce to the equations
\begin{align}\label{eq:poincaredualgiant}
\partial_1 B_1-\cos\theta \partial_2 B_3+\partial_3
B_3&=0\,,\nonumber\\
\sin\theta (\partial_1(\sin \theta B_3))-\partial_3
B_1+\cos\theta \partial_2 B_1&=0\,,
\end{align}
where all fields are time-independent as a result of the remaining Bianchi identities. Notice that the gauge field configuration on this kind of brane is like a `snapshot' of the propagating wave found on the dual giant in global AdS$_2$ coordinates above.
Defining $G=B_1 + i\sin\theta B_3$ and taking linear combination again yields (\ref{eq:complex}).
The mechanical momentum densities are
\begin{align}
{\cal P}_{\xi_i} &=  T_3\left(\frac{l}{4\omega }\right)^2\csc\theta\left[ 16|G|^2 + \omega^4\sin^2\theta \right]\mu_i^2 \,,
\end{align}
while the field gives rise to 
\begin{align}
{\cal P}_{E_1} = \frac{\partial {\cal L}}{\partial E_1} = T_3 B_3 \sin\theta\,,\quad{\cal P}_{E_2} = \frac{\partial {\cal L}}{\partial E_2} = T_3 B_1 \cot\theta\,, \quad{\cal P}_{E_3} = \frac{\partial {\cal L}}{\partial E_3} = -T_3 B_1 \csc\theta\,.
\end{align}
As their un-fluxed counterparts, these configurations satisfy the BPS relation $
{\cal H}=0$. Let us now turn to giant-like configurations, \textit{i.e.} configuration that wrap a submanifold in the S$^5$ part of the geometry.

\subsubsection{Global $\BHG$}\label{sec:globalgiants}
Let us consider a giant-like configuration with world-volume
coordinates $\mathbf{\sigma}=\{
\tau\,,\beta\,,\xi_2\,,\xi_3 \}$.
We want to put a non-trivial gauge field configuration on the solution in \cite{sss} with
\begin{equation}
\dot\psi=0,\quad \dot\phi=-\frac{2l}{\omega\lambda},\quad \dot\xi_1=-\frac{2\omega}{l\lambda}\,\quad \dot\alpha=0\,\quad\dot\rho=0
\end{equation}
 The supersymmetry constraints are solved by the relations
\begin{equation}
E_2=-E_3=-{2\omega\over l\lambda} B_1\,,\quad B_3
  =-B_2\tan^2\beta\,,\quad E_1\cos^2\beta={2\omega\over l\lambda} B_2
  \,.
\end{equation}
The fact that $F\wedge F$ vanishes is again telling us that the electric
  and magnetic fields are perpendicular to one another. With these solutions, we see that again
\begin{equation*}
\sqrt{-\det(h+F)}=\sqrt{-\det(h)}\,,
\end{equation*}
so that the same linear combination of supercharges is preserved with flux, as without flux. 
Thus, indeed we have an EM-wave running in the directions $\xi_2$ and
$\xi_3$. The Bianchi identities and coordinate equations of motion
determine the time dependence of the waves to be
\be
\partial_3 E_i + \partial_2 E_i = \frac{l\lambda}{2\omega}\partial_0
E_i\,,\quad i=2,3\,.
\ee
\vskip.5em
{\it The phase velocity of the waves, $\frac{2\omega}{l\lambda}$, is again exactly as expected from the
  BPS condition.}
\vskip.5em
The gauge field equations of motion (\ref{eq:gaugefield}) together with the Bianchi identities on the configuration under consideration here lead to the system of equations
\bea\label{gcons}
\sin 2 \beta \partial_1 (\sin 2 \beta
E_1)-2\partial_3 E_2+2\partial_2 E_2+\frac{l\lambda}{2\omega}\cos 2 \beta \partial_0 E_2&=&
0\,,\nonumber\\
\partial_2 E_1-\partial_3 E_1-2 \partial_1 E_2 +\frac{l\lambda}{ \omega}\cos 2 \beta \partial_0 E_1&=& 0\,.
\eea
Upon identifying $2\beta\rightarrow \theta$,
$\xi_2-\xi_3\rightarrow\psi$ and $\xi_2 + \xi_3\rightarrow\phi$ we can
recast this computation into the standard form above. The quantity
$G=2E_2 + i\sin 2 \beta E_1$ satisfies (\ref{eq:complex}). 
The mechanical momentum densities pick up contributions due to the field:
\begin{align}
{\cal P}_{\phi} & = -T_3\frac{l^2}{24\omega^2}\frac{1}{\sin 2\beta
  \cos^2\alpha}\left(  \lambda^2 |G|^2 + \omega^2 \cos^4\alpha \sin^2
  2\beta (4l^2 + 3 \omega^2 \cos^2\alpha)\right)\,,\\
{\cal P}_{\psi} & = \cos\theta {\cal P}_{\phi}\,,\\
{\cal P}_{\xi_1} &= - T_3\frac{l^2}{8\omega^2}\frac{\tan^2\alpha}{\sin 2\beta}\left( \lambda^2 |G|^2 + 4 l^2\omega^2\cos^4\alpha\sin^2 2\beta \right)\,.
\end{align}
Note that in the absence of $G$, there was an upper bound in the
momenta. Since $\alpha$ runs between $0$ and $\pi/2$ there is no such
upper bound any more. This seems to hint at the interpretation of
fluxes on giants as descendants \cite{bglm}. However since AdS$_5$ dual
giants with fluxes are not analogous to their S$^5$ counterparts, the
corresponding interpretation of fluxes on dual giants as descendants
is less clear. 
The gauge field degrees of freedom have conjugate momentum densities
\begin{align}
{\cal P}_{E_1} = \frac{\partial {\cal L}}{\partial E_1}=T_3
B_2\tan\beta\,,\quad {\cal P}_{E_2} =
\frac{\partial {\cal L}}{\partial E_2}= -T_3
 B_1\cot\beta\,,\quad {\cal P}_{E_3} =
\frac{\partial {\cal L}}{\partial E_3}= T_3 B_1\tan\beta\,.
\end{align}
The Hamiltonian density gives rise to the BPS relation
\begin{equation}
{\cal H} = \frac{2l}{\omega\lambda}|{\cal P}_\phi| + \frac{2\omega}{l\lambda}\sum_{i=1}^3|{\cal P}_{\xi_i}|\,,
\end{equation}
where
\begin{equation}
{\cal P}_{\xi_2}= B_3{\cal P}_{E_1}-B_1 {\cal
  P}_{E_3}\,,\quad {\cal P}_{\xi_3} = B_1 {\cal P}_{E_2} - B_2 {\cal P}_{E_3}\,.
\end{equation}
\subsubsection{Poincar\' e $\BHG$}
Solving the supersymmetry constraints for $D3$ branes wrapping
$\left\{t,\beta,\xi_2,\xi_3\right\}$ in AdS$_2$ Poincar\'e
coordinates, where $\dot\xi_1 = \dot\theta=\dot\phi=\dot\psi=\dot r=0$
, results in the constraints
\begin{equation}
E_i=0,\qquad B_3 + \tan^2\beta \,B_2=0
\end{equation}
with $B_1$ unconstrained. Taking note of fact that $\gamma_0
\epsilon^*=0$ (see \cite{sss}) simplifies the calculation. The DBI part of the action on this class of solutions again simplifies in the same way as above. The equations of motion for an anti-brane and associated Bianchi identities reduce to
\begin{align}
\partial_{2}B_2 -\cot^2\beta \partial_{3}B_2 + \partial_1 B_1 &=0\,,\nonumber\\
\partial_1 (\tan\beta B_2) + \partial_{3}\tan\beta B_1 - \partial_{2}\cot\beta B_1&=0\,
\end{align}
with all magnetic field components time-independent. Defining the auxiliary variables 
\begin{equation*}
B_2 = \frac{l\lambda}{2\omega}\cos^2\beta G_1\,,\qquad G_2 = -G_3 = -\frac{2\omega}{l\lambda}B_1\,.
\end{equation*}
and identifying   $2\beta \rightarrow \theta$, $\xi_2 -
\xi_3\rightarrow \psi$ and $\xi_2 + \xi_3\rightarrow \phi$, after some algebra, transforms
the equations  into standard  form (\ref{eq:complex}) in terms of the complex field
 $G=2 G_3 + i\sin 2\beta G_1$.
 The mechanical momentum densities pick up contributions due to the field:
\begin{align}
{\cal P}_{\phi} & = T_3{1\over 8 \sin 2 \beta \cos^2\alpha }(\lambda^2 |G|^2+ l^2 \omega^2
\cos^4\alpha\sin^2 2 \beta (4-\cos^2\alpha))\,,\\
{\cal P}_{\psi} & = \cos\theta {\cal P}_{\phi}\,,\\
{\cal P}_{\xi_1} &=  T_3\frac{l^2}{8\omega^2}\frac{\tan^2\alpha}{\sin 2\beta}(\lambda^2 |G|^2+4 l^2 \omega^2 \cos^4\alpha \sin^2 2\beta)\,.
\end{align}
As in the global case, the ${\cal P}_i$ do not have upper limits
any more.
The gauge field degrees of freedom have conjugate momentum densities
\begin{align}
{\cal P}_{E_1} = \frac{\partial {\cal L}}{\partial E_1}=T_3
B_2\tan\beta\,,\quad {\cal P}_{E_2} =
\frac{\partial {\cal L}}{\partial E_2}= -T_3
 B_1\cot\beta\,,\quad {\cal P}_{E_3} =
\frac{\partial {\cal L}}{\partial E_3}= T_3 B_1\tan\beta\,.
\end{align}
  The solutions satisfy the BPS relation ${\cal H}=0$ in terms of their conjugate momenta as expected from the Killing spinor bilinear.
\section{Discussion}
In this paper we discussed microstates in the 
near-horizon geometry of a $\frac{1}{16}$-BPS AdS$_5$ black hole. We counted dual
giant configurations in the probe approximation by quantising the
phase space of solutions. The result missed the macroscopic entropy by
a degeneracy factor. We argued that turning on an additional
angular momentum quantum number, achieved in this paper by world-volume fluxes and
dictated by the near-horizon supersymmetry, can potentially produce
the correct statistical entropy.

We found a whole class of solutions preserving exactly
the same supersymmetry as those without fluxes. These solutions are
BPS electromagnetic waves and are entirely consistent with the
supersymmetries of the near-horizon geometry. They have precisely the
velocity predicted by supersymmetry and exist on the world volumes of
both giants and dual giants. The resulting configurations carry all
four quantum numbers dictated by supersymmetry. We also demonstrated that world-volume fluxes on $\frac{1}{8}$-BPS dual giants
in AdS$_5\times$ S$^5$ will generically lead to $\frac{1}{16}$-BPS configurations
with an additional quantum number. It will be interesting to consider
the partition functions of these states along the lines of \cite{ms}.

The global $\tBHG$ configurations in this paper may be viewed as the
caps of the microstates of the full black hole in the fuzzball
\cite{Mathur:2005zp} 
proposal\footnote{We thank N. Suryanarayana for suggesting this to us.}.
It will be very interesting to consider the quantisation of the new space
of near-horizon solutions and to see if the macroscopic entropy is
reproduced. A simple minded maximisation argument was shown to lead to an exact
match with the macroscopic entropy although we should emphasise that
this cannot be construed as a satisfactory derivation as yet. What was
crucial in this argument was the existence of the additional angular
momentum quantum number, one source of which are the electromagnetic
waves. There could be other sources such as vibrational modes which we
have not ruled out. A related
puzzle is that fluxes on $\frac{1}{8}$-BPS AdS$_5\times$S$^5$ giants are to be thought of as
descendants \cite{bglm} and as such would lead to double counting. If
the same interpretation extends to $\frac{1}{16}$-BPS AdS$_5$ dual giants with
fluxes, then these should be thought of as descendants of some chiral
primary operators presumably corresponding to BPS vibrational
modes \cite{Das:2000st}. However, since electromagnetic waves broke supersymmetry in the
dual giant case, that this analogy holds is not clear to
us.  
Although it is expected that BPS fluctuations or mechanical waves also
should play a role in the counting of microstates, it is not
implausible that the electromagnetic waves are just a dual description
of these mechanical waves. Electromagnetic flux is related
to open strings while the vibrational modes are related to the metric, so
this would be similar in spirit to open-closed duality. Since it
appears that finding solutions for mechanical waves is considerably harder, one
could hope that counting the electromagnetic waves in a systematic way
could reproduce the same result. Our analysis
should be useful in studying these issues further.

\acknowledgments{We thank Nemani Suryanarayana for collaboration at an
  initial stage of this work, for useful discussions and comments on
  the manuscript. We would also like to
  thank Diego Correa, Nick Dorey, Jan Gutowski, Gustav Holzegel, David Tong and Claude Warnick for 
  discussions. AS is supported by PPARC and Gonville and Caius
  college, Cambridge. JS thanks the Gates Cambridge Trust and PPARC for financial support and the EFI at the University of Chicago for hospitality during the final stages of preparation of this paper.}
\appendix
\section{Details of Computations for Dual Giants}
In this appendix we supply more detail on the supersymmetry analysis whose results were quoted in section \ref{sec:globaldualgiants}. On the dual giant wrapping $\{\tau,\theta,\phi,\psi  \}$ in AdS$_2$ global coordinates, we have the induced metric
\begin{equation}
h=\begin{pmatrix} -1+J^2 l^2 & 0 & -{J\omega^2\over 4} & -{1\over 4}
J\omega^2\cos\sigma_1\\
0 & {\omega^2\over 4} & 0 & 0\\
-{J\omega^2\over 4} & 0 & {\omega^2(l^2+\omega^2)\over 4 l^2} &
{\omega^2(l^2+\omega^2)\over 4 l^2} \cos\sigma_1\\
-{1\over 4} J\omega^2\cos\sigma_1 & 0 & {\omega^2(l^2+\omega^2)\over 4
  l^2} \cos\sigma_1 & {\omega^2 (2 l^2+\omega^2+\omega^2 \cos
  2\sigma_1)\over 8 l^2}
\end{pmatrix}
\end{equation}
where $J=\mp\frac{2\omega}{l\lambda}$ for a brane /anti brane. We wish to preserve the same linear combinations of supercharges as in the unfluxed case, discussed in \cite{sss}. The preserved Killing spinor satisfies the relation
\begin{equation*}
(h_{02}-\gamma_0\gamma_2)\epsilon=-b\epsilon\,,
\end{equation*}
so that
\begin{equation}
\gamma_0\epsilon^*=-\frac{b+h_{02}}{h_{22}}\gamma_2\epsilon^*\,.
\end{equation}
Using this we find that the vanishing of the $\epsilon^{ijkl}F_{ij}\gamma_{kl}$ term in the $\kappa$ symmetry projector sets
\begin{eqnarray}\label{const}
\Biggl[  E_1(-\gamma_3\gamma_2+h_{23})+E_2 \gamma_{1}\gamma_3+&E_3&
\gamma_1\gamma_2+B_3\left(\frac{b+h_{02}}{h_{22}}\gamma_{3}\gamma_2
+h_{03}\right)\nonumber +\nonumber \\
 &&\left(b-\frac{J \omega^2}{2}\right) B_2-B_1 \frac{b+h_{02}}{h_{22}}\gamma_1\gamma_2\Biggr]\epsilon^*=0\,.
\end{eqnarray}
%
%
Using the explicit form of the world-volume gamma matrices, we compute the various terms appearing above.
\bea
-\gamma_3 \gamma_2+ h_{23} &=& \frac{\omega^2}{4l}\sin\theta
(\sin\phi\Gamma_3-\cos\phi \Gamma_2)\left(\frac{\omega}{2}\Gamma_9+\frac{l}{ab}\Gamma_4\right)\,,\nonumber\\
\frac{b+h_{02}}{h_{22}} \gamma_3\gamma_2+h_{03}&=&
\left(b-{J\omega^2\over 2}\right)\cos\theta \nonumber\\
&&+\frac{b-\frac{J\omega^2}{ 4}}{h_{22}}\frac{\omega^2}{4l}\sin\theta (\sin\phi\Gamma_3-\cos\phi\Gamma_2)\left(\frac{\omega}{2}\Gamma_9+\frac{l}{ab}\Gamma_4\right)\,,\nonumber \\
\gamma_1\gamma_3&=& \frac{\omega^2}{4l}(\cos\phi\Gamma_3+\sin\phi\Gamma_2)\cos\theta\left(\frac{\omega}{2}\Gamma_9+{l\over ab}\Gamma_4\right)+{\omega^2\over
  4}\sin\theta\Gamma_{23}\,,\nonumber\\
\gamma_1\gamma_2 &=& \frac{\omega^2}{4 l}(\cos\phi
\Gamma_3+\sin\phi\Gamma_2)\left(\frac{\omega}{2}\Gamma_9+\frac{l}{ab}\Gamma_4\right)\,.
\eea
One now projects the resulting equations onto the subspace defined by
the projection conditions for $\tBHG$ and equations independent
generators of the Clifford algebra to zero individually. The constant
term in equation  (\ref{const}) gives
\begin{equation}
\left(b-{J\omega^2\over 2}\right) (B_3\cos\theta+B_2)-i{\omega^2\over 4}\sin\theta E_2\,,
\end{equation}
demanding this to be zero gives
\be
E_2=0\,,\quad B_3 \cos\theta=-B_2\,.
\ee
Further the coefficient (after using the projection condition
$\Gamma_{23}\epsilon=i\epsilon$ ) of $e^{i\phi}\Gamma_2({\omega\over
  2}\Gamma_9+{l\over ab}\Gamma_4)$ gives
\bea
E_1&=& -B_3\left(b-\frac{J\omega^2}{4}\right)\frac{1}{h_{22}}=-\frac{2l}{\omega \lambda} B_3\,,\\
E_3&=& B_1 \left(b-\frac{J\omega^2}{4}\right)\frac{1}{h_{22}}=\frac{2l}{\omega
  \lambda} B_1\,.
\eea
These are the relations used in section
\ref{sec:globaldualgiants}. The Poinar\'e computation goes ahead in
much the same way, but is algebraically simpler.
\section{Details of Computations for Giants}
For a brane wrapping $\{\tau, \beta,\xi_2,\xi_3 \}$ we have the induced metric
\begin{equation}
h_{\rm giant}=\left(  \begin{array}{cccc}
\frac{2\omega^2}{\lambda^2}(\mu_1^2 -1) & 0 &\frac{l\omega}{2\lambda}\mu_2^2 & \frac{l\omega}{2\lambda}\mu_3^2\\
0&l^2 (1-\mu_1^2) & 0 & 0\\
\frac{l\omega}{2\lambda} \mu_2^2 & 0 & l^2 \mu_2^2 & 0\\
\frac{l\omega}{2\lambda}\mu_3^2 & 0 & 0 & l^2 \mu_3^2
                      \end{array}\right)\,.
\end{equation}
We compute
\begin{align}
\gamma_1 \gamma_2 & = \phantom{-}l^2 \mu_2 \left[ \mu_3 \Gamma_{68} - \mu_2 \left( \sqrt{1-\mu_1^2}\,\Gamma_{69} + \mu_1 \Gamma_{67} \right) \right]\nonumber\\
\gamma_1 \gamma_3 &=-l^2 \mu_3 \left[   \mu_2 \Gamma_{68} + \mu_3 \left( \sqrt{1-\mu_1^2}\,\Gamma_{69} + \mu_1 \Gamma_{67} \right) \right]\nonumber\\
\gamma_2\gamma_3 &=-l^2 \mu_2\mu_3 \left[\sqrt{1-\mu_1^2}\,\Gamma_{89} + \mu_1 \Gamma_{87} \right]\nonumber\,.\\
\end{align}
The supersymmetry constraint (\ref{eq:fluxsusycondition}) now reads
\begin{eqnarray}
\left[ E_1 \gamma_2\gamma_3  -E_2 \gamma_1\gamma_3 + E_3\gamma_1\gamma_2  + B_3 \left(-\gamma_3\gamma_0 + \frac{l\omega}{2\lambda}\mu_3^2 \right) \right.\nonumber\\
\left. + B_2 \left( - \gamma_2\gamma_0 + \frac{l\omega}{2\lambda}\mu_2^2  \right) - B_1\gamma_1\gamma_0 
\right] \epsilon^*=0\,.
\end{eqnarray}
Since we want to preserve the same supersymmetries as without flux, we can use the fact that
\begin{equation}
\Gamma\epsilon^*=-\frac{1}{\sqrt{-\det h}}\left( \gamma_0\gamma_1\gamma_2\gamma_3 - h_{03}\gamma_1\gamma_2 + h_{02}\gamma_1\gamma_3 \right)\epsilon^*=i\epsilon^*
\end{equation}
for the case of a brane. Plugging in the expressions for $\gamma_1\gamma_2$ and $\gamma_1\gamma_3$ from above, we arrive at the equation
\begin{align}
\gamma_0\gamma_1\gamma_2\gamma_3 \epsilon^* &= -i\left( \sqrt{-h} - \frac{l^3\omega}{2\lambda}\mu_2\mu_3(1-\mu_1^2) \right)\epsilon^*\\
&\equiv -iA\epsilon^*\,.
\end{align}
It is easy to compute
\begin{equation}
A = -\frac{l^2 \lambda}{2\omega}\mu_2\mu_3 h_{00}\,.
\end{equation}
Now we may write the supersymmetry condition
\begin{eqnarray}
\left[\left(E_1 - \frac{i}{A}h_{00}h_{11}B_1  \right) \gamma_2\gamma_3 -\left(E_2 - \frac{i}{A}h_{00}h_{22}B_2  \right) \gamma_1\gamma_3 \right.\\ +\left(E_3 - \frac{i}{A}h_{00}h_{33}B_3  \right) \gamma_1\gamma_2 +		
\left. \frac{l\omega}{2\lambda}\left(  B_3\mu_3^2 + B_2\mu_2^2\right) \right]\epsilon^*=0\,.
\end{eqnarray}
From this we may extract the coefficient equations of $\left\{ \mathbb{1},\Gamma_{69},\Gamma_{67} \right\}$. Start with the coefficient of the unit matrix:
\begin{equation}
i l^2\mu_2\mu_3 (E_2+E_3) + \frac{l^2 h_{00}}{A}\mu_2\mu_3\left(  h_{22}B_2 + h_{33}B_3\right) + \frac{l\omega}{2\lambda}\left( B_3\mu_3^2 + B_2\mu_2^2  \right)=0\,.
\end{equation}
We must equate real and imaginary parts to zero individually and obtain
\begin{equation}
E_2 = -E_3\,\qquad B_3\cos^2\beta = - B_2\sin^2\beta\,.
\end{equation}
Next, we turn to the coefficient of $\Gamma_{67}$:
\begin{equation}
-i(E_1-{i\over A}h_{00}h_{11}B_1)\mu_2\mu_3+\mu_3^2(E_2-{i\over
  A}h_{00}h_{22}B_2)-\mu_2^2(E_3-{i\over A}h_{00}h_{33}B_3)=0\,,
\end{equation}
while the coefficient of $\Gamma_{69}$ gives
\begin{equation}
-i(E_1-{i\over A}h_{00}h_{11}B_1)\mu_2\mu_3+\mu_3^2(E_2-{i\over
  A}h_{00}h_{22}B_2)-\mu_2^2(E_3-{i\over A}h_{00}h_{33}B_3)=0\,,
\end{equation}
which are the same conditions. Equating the real and imaginary parts
  gives
\begin{equation}
-{2\omega\over l \lambda} B_1=E_2\,,\quad E_1={2\omega \over l
  \lambda}(B_2-B_3)\,.
\end{equation}
 A  cross-check that these are correct is to compute
\begin{equation}
\epsilon^{\mu\nu\rho\lambda}F_{\mu\nu}F_{\rho\lambda}=E_1 B_1+E_2
  B_2+E_3 B_3=0\,,
\end{equation}
which is needed for susy to hold. Thus the most general solution is:
\begin{equation}
E_2=-E_3=-{2\omega\over l\lambda} B_1\,,\quad B_3
  =-B_2\tan^2\beta\,,\quad E_1\cos^2\beta={2\omega\over l\lambda} B_2
  \,.
\end{equation}
This is the set of constraints used in section
\ref{sec:globalgiants}. Again, the Poincar\'e case proceeds analgously.

\end{document}